\documentclass[final]{aipproc}
\layoutstyle{8x11single}
\begin{document}

\title{Constraining Quark Transversity through Collins Asymmetry Measurements at STAR}

\classification{13.85.Ni,13.87.Fh,13.88.+e}
\keywords      {RHIC,STAR,jets, transversity}

\author{Renee Fatemi, for the STAR Collaboration}{
  address={University of Kentucky, Lexington, KY, 40506}
}

\begin{abstract}
The quark transversity distributions are accessible via measurements of the azimuthal distribution of charged pions inside jets produced in collisions of transversely polarized protons.  The STAR Detector at the Relativistic Heavy Ion Collider is capable of full jet reconstruction and charged pion identification in the mid-rapidity region. This proceeding presents the first results of the Collins moment of leading charged pions in jets reconstructed from 2.2 $pb^{-1}$ of $\sqrt{s}=200$ GeV transversely polarized (58$\%$) proton data. 
 \end{abstract}

\maketitle

\section{Motivation}

Investigations into the partonic spin structure of the proton have repeatedly led to unexpected results.  The first measurement of the total quark polarization in longitudinally polarized protons was much smaller than expected\cite{Ashman:1989ig} and spawned decades of experiments\cite{deFlorian:2009vb} that now constrain the quark contribution to be $\sim30\%$.  Similarly, the left-right asymmetries of pions produced in transversely polarized proton collisions were found to be several orders of magnitude larger\cite{Adams:1991rw}\cite{Adams:1991cs} than estimated from initial perturbative Quantum Chromodynamics (pQCD) calculations.  This discovery encouraged physicists to expand the traditional collinear framework used to define and extract the spin dependent parton distribution functions.  The introduction of distribution functions that depend on the transverse momentum  of the partons ($k_T$) inside the proton and fragmentation functions that depend on the transverse momentum of hadronization products of the partons ($j_T$)  implies that the proton has a much richer spin dependent substructure than originally envisioned. \\

\noindent 
Ironically, the introduction of $j_T$ dependent fragmentation functions, referred to as the Collins Functions\cite{Collins:1992kk} ($\Delta{D(z)}$),  in turn facilitated the experimental measurement of the least known distribution function within the $k_T$ integrated framework.  The transversity distribution \cite{Ralston:1979ys} characterizes the number density of transversely polarized partons inside a transversely polarized nucleon.  The momentum ($f(x)$),  helicity ($\Delta{f}(x)$) and transversity  ($\Delta_T{f(x)}$) distributions completely describe the momentum and spin structure of the proton at leading twist within a collinear framework\cite{Jaffe:1991kp}.  Unlike $f(x)$ and $\Delta{f(x)}$,   $\Delta_Tf(x)$ is chiral odd and can be observed only when coupled with another chiral odd operator, such as $\Delta{D(z)}$.  Measurements in semi-inclusive deep inelastic (SIDIS) and electron-positron scattering (see references in \cite{Anselmino:2009zz}) have shown $\Delta{D(z)}$ to be sizable and have provided input to the first global analysis of the transversity and Collins distributions\cite{Anselmino:2009zz}\cite{Anselmino:2007fs}\cite{Bacchetta:2011ip}.  

\section{Accessing Transversity in Hadronic Collisions}

Yuan\cite{Yuan:2007nd} was the first to propose that the azimuthal distribution of charged pions within a jet in hadronic collisions would provide sensitivity to the $\Delta_T{f(x)}\otimes\Delta{D(z)}$ distributions.  A quark from a transversely polarized proton beam, carrying momentum fraction $x=p_z^{quark}/p_z^{proton}$, scatters off a parton from an unpolarized proton beam.  Global analyses have shown that $\Delta_T{f(x)}$ is non-zero so the quark from the polarized proton will carry a spin orientation preference associated with the proton spin.  When the quark fragments,  the emerging hadron's $j_T$  is correlated with the spin direction of the quark.  This combination of effects will produce an asymmetric distribution of pions within a reconstructed jet that depends on the initial spin orientation of the parent proton.\\

\begin{figure}[h]
  \includegraphics[height=0.2\textheight]{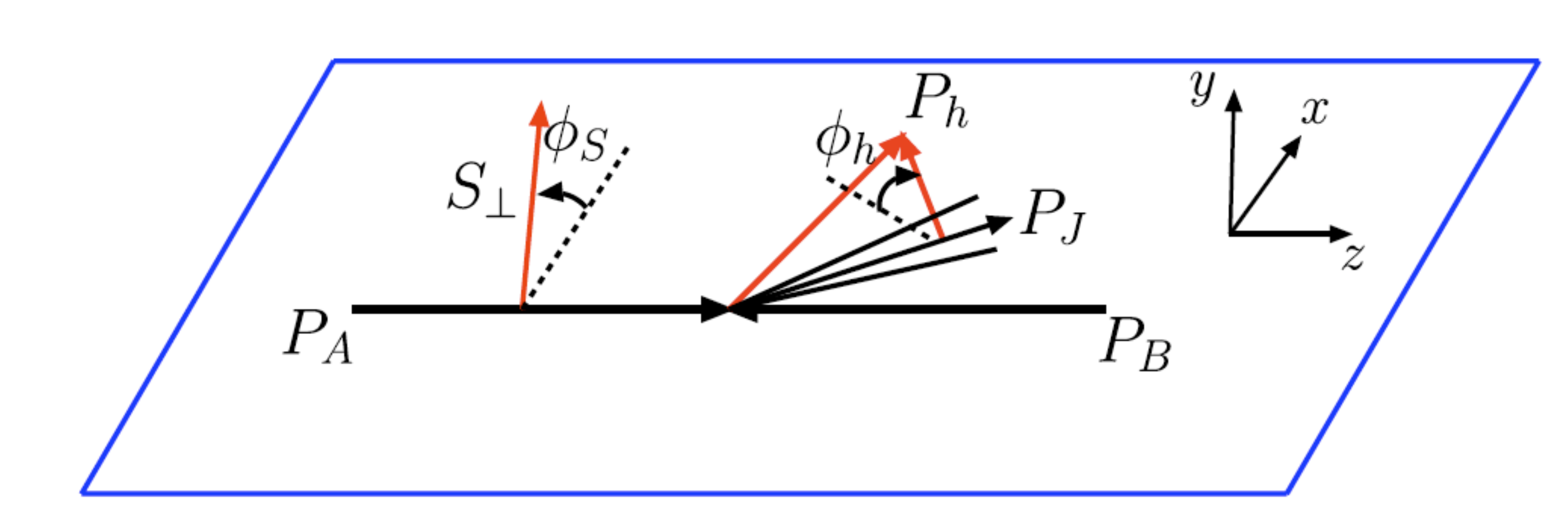}
  \label{fig:kinematics}
\caption{Kinematic diagram of the reaction plane, $\phi_S$ and $\phi_H$ as defined in \cite{Yuan:2007nd}}
\end{figure}

\noindent The differential cross-section for this process can be written in terms of the unpolarized cross-section and a sine weighted moment associated with the Collins function:

\begin{equation}
\frac{d\sigma}{d\Omega} = \frac{d\sigma_{unpol}}{d\Omega}(1 + A_N\sin(\phi_H- \phi_S))
\label{eq:collinsmoment}
\end{equation}

 \noindent As shown in figure \ref{fig:kinematics}, $\phi_S$ is the angle of the polarized proton spin direction with respect to the reaction plane formed by the reconstructed jet and the beam axis.  $\phi_H$ is the angle between the transverse momentum vector of the charged pion with respect to the jet axis and the reaction plane.  $A_N$ is related to $\Delta_T{f(x)}\otimes\Delta{D(z)}$.  Yuan's treatment has been expanded by D'Alesio, Murgia and Pisano\cite{D'Alesio:2010am} to include $k_T$ dependent distributions in the initial state.  This generalized approach indicates the Collins moment is also sensitive to contributions from terms involving the convolution of Sivers and Boer-Mulders distributions in addition to the $\Delta_T{f(x)}$ term.  D'Alesio \textit{et al.} have estimated this additional term to be negligible even in maximized scenarios. Data from high $p_T$ jets in hadronic collisions are complementary to SIDIS measurements, providing crucial tests of the universality of the Collins functions and information on the scale evolution of $\Delta_T{f(x)}$.

\section{Experimental Measurement}
STAR is a large acceptance detector\cite{Ackermann:2002ad} located at the Relativistic Heavy Ion Collider (RHIC).  The Time Projection Chamber (TPC) performs track trajectory and momentum reconstruction, as well as particle identification for charged particles scattered at mid-rapidity  ($|\eta|<1.4$).  The Electromagnetic Calorimeters (EMC) provide triggering capabilities as well as  electromagnetic energy and position information for  $-1<\eta<2$.  Segmented Beam-Beam Counters (BBC) located up and downstream of the STAR interaction region measure the spin dependent relative luminosities and provide the minimum bias trigger condition.  The jet patch trigger (JP) used in this analysis required $E_T > 7.8$ GeV to be deposited within a patch of $\Delta\eta\times\Delta\phi=1\times1$ in the EMC.  An integrated luminosity of 2.2 pb$^{-1}$ from the 2006 JP trigger data was used in this analysis. RHIC provided collisions of two transversely polarized proton beams of $\sqrt{s} =200$ GeV and average beam polarization of $\sim58\%$.  The analysis is performed separately for each beam (blue and yellow), treating one beam as polarized while integrating over the spin of the other beam.  The final result combines the two beam analyses. \\

\begin{figure}[h]
 \begin{minipage}{8.0cm}
  \includegraphics[height=.22\textheight]{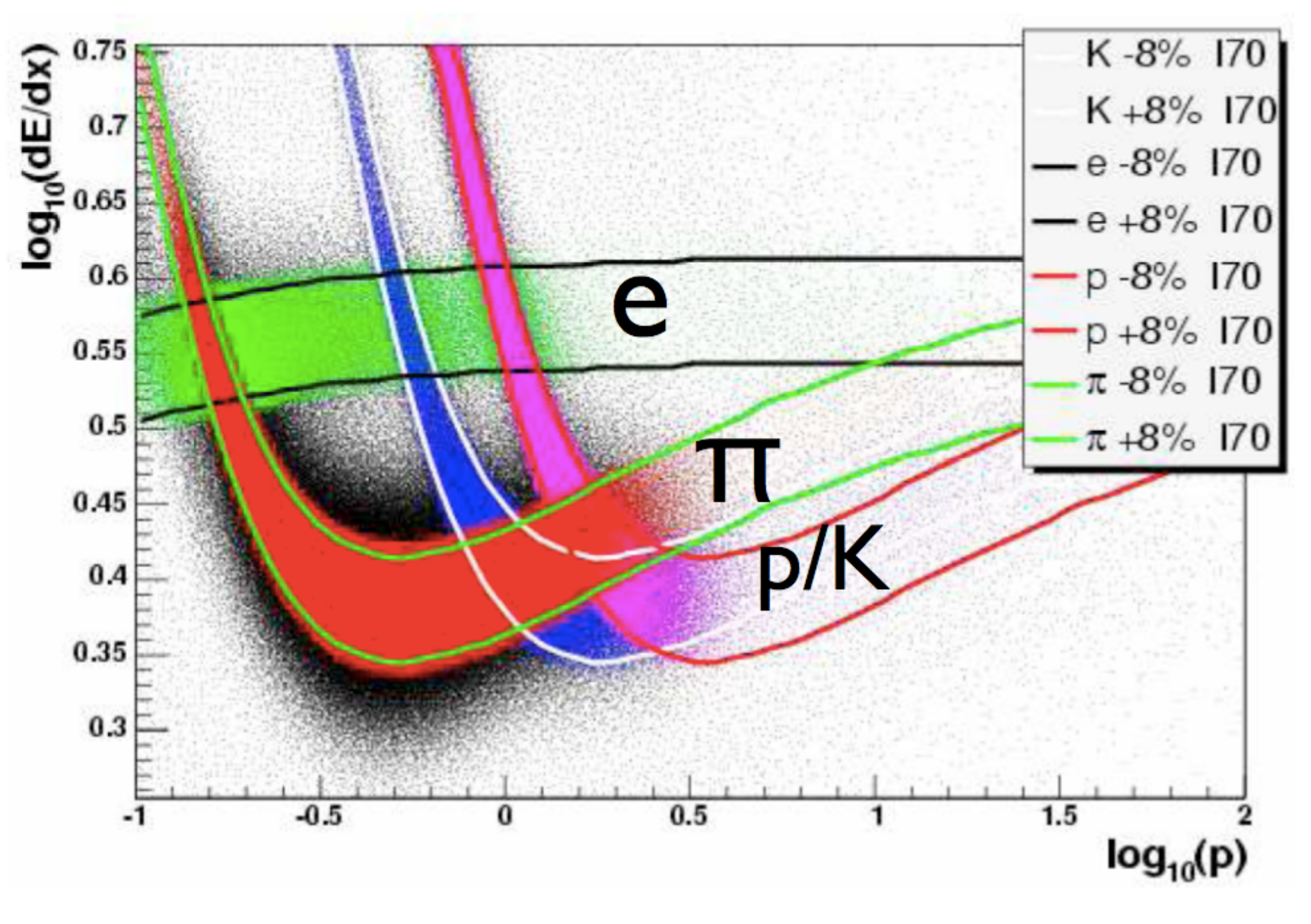}
\end{minipage}
\caption{The left plot shows the distribution of $\log_{10}(dE/dx)$ as a function of $\log_{10}(p)$ for electrons, pions,
kaons and protons. The color bands indicate the $\pm{1}\sigma$ $dE/dx$ resolution\cite{Shao:2005iu}.   The right plot shows the $n\sigma(\pi)$ distribution for all particles identified as leading $\pi^-$.  The red, blue  and green curves represent the contributions to the total sample (black) from $\pi^-$, electrons and kaons+protons.}
\hfill
\begin{minipage}{6.0cm}
  \includegraphics[height=.2\textheight]{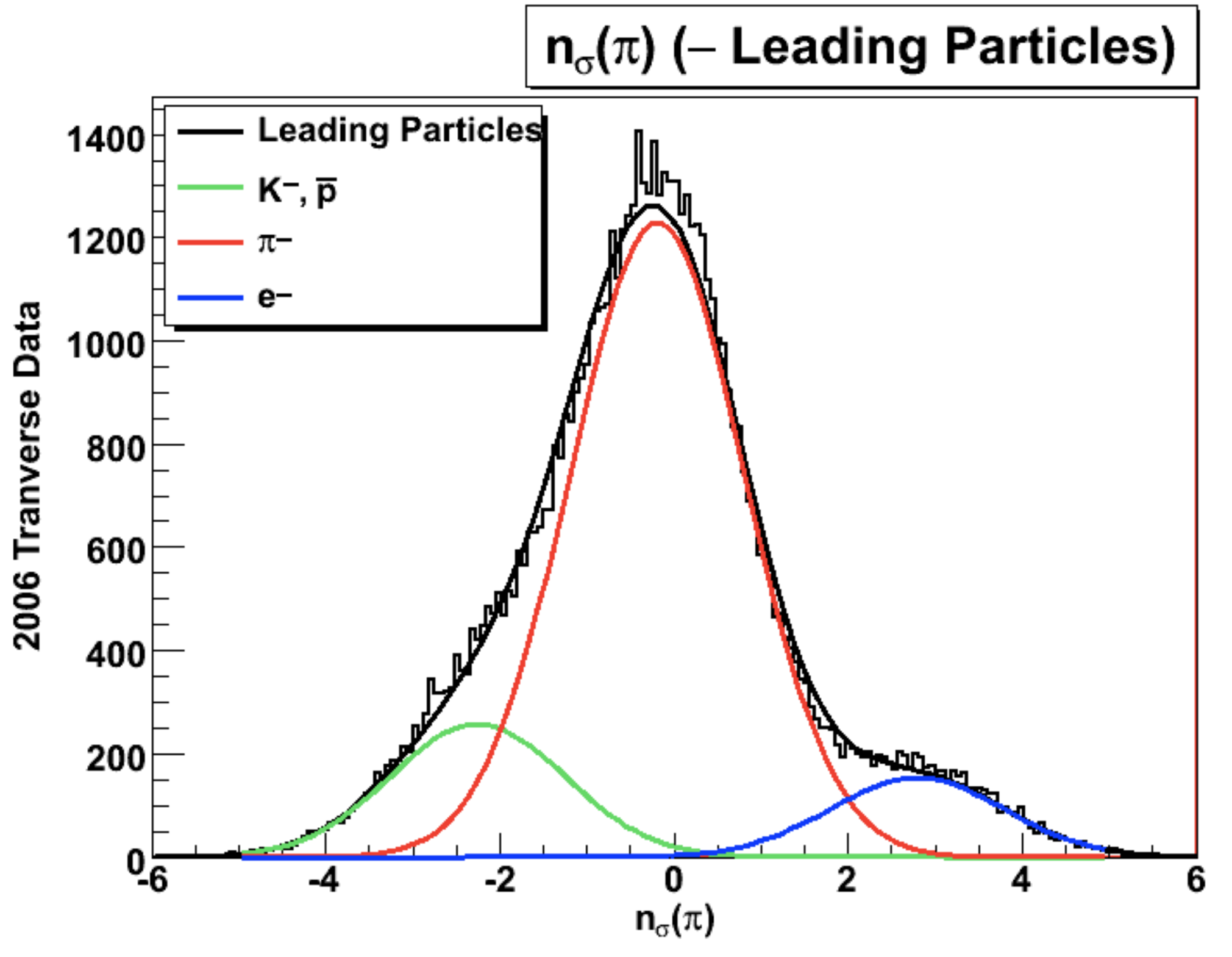}
\end{minipage}
\hfill   
\label{fig:pions}
\end{figure}

\noindent Jets were reconstructed using the mid-point cone algorithm\cite{Blazey:2000qt} with radius R = 0.7 and split/merge fraction of 0.5.  In an effort to enrich the sample with jets originating from the transversely polarized beam under consideration, only jets in the forward region (with respect to the beam) were considered.  This translates into a cut of $0<\eta<0.9$ $(0>\eta>-0.9)$ for the blue (yellow) beam.  The fraction of gluon and quark jets resulting from proton scattering varies with jet $p_T$,  with gluon jets dominating below 10 GeV and quark jets above  20 GeV.  Since the gluon transversity distribution in the proton must be zero at leading twist,  only jets with $p_T$  >  10 GeV are analyzed in order to reduce the dilution from the gluon signal to the measured asymmetry.  Charged pions were identified by placing an asymmetric cut ($-1<n{\sigma(\pi)}<2$) on the characteristic $dE/dx$ curve in the TPC, shown in figure \ref{fig:pions}.  The $n\sigma(\pi)$ distribution is the $\log(dE/dx)$ distribution divided by the expected mean and $dE/dx$ resolution for pions.  The cut was optimized to minimize contamination by electrons and kaons, currently estimated to be < 6$\%$. \\

\noindent The measured Collins moment was calculated by forming the ratio of the of sum of the $\sin(\phi_H - \phi_S)$ weighted events and the sum of the polarization ($P$) weighted events:

\begin{equation}
A_{meas}(z,j_T) = \frac{2\sum\limits_{\phi_C}{N(z,j_T)\sin(\phi_C)}}{\sum\limits_{\phi_C}{PN(z,j_T)}}= A_N(z,j_T)
\end{equation}

\noindent Here  $\sin(\phi_C)=\sin(\phi_H - \phi_S)$ and $A_N$ is the asymmetry defined in eq. \ref{eq:collinsmoment}.  Events were also normalized by the spin dependent luminosities according to the polarized beam spin orientation. The measurement is binned in both pion momentum fraction $z$ and pion $j_T$.  Figure \ref{fig:results} shows the preliminary results from STAR, for both positively and negatively charged pions, plotted as a function of $z$ ($j_T$) and integrated over $j_T$ ($z$). \\

\begin{figure}[h]
 \label{fig:results}
  \includegraphics[height=.3\textheight]{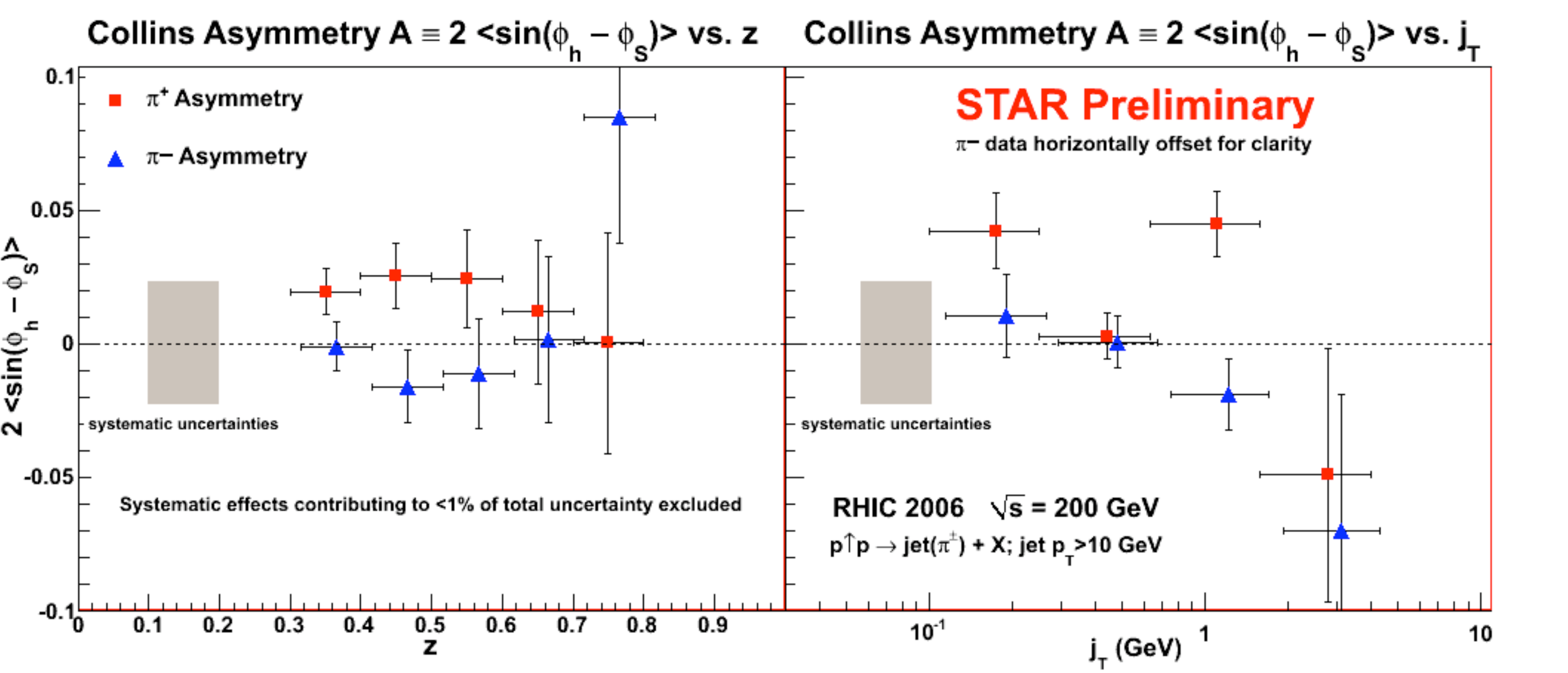}
  \caption{Preliminary results of the Collins moment for leading $\pi^+$(red) and $\pi^-$(blue) particles within mid-rapidy jets reconstructed by the STAR detector.  Statistical errors are shown on data points and the grey shaded band indicates the systematic error bar for the $\pi^+$  and $\pi^-$ separately.  The momentum fraction of the jet carried by the pion is $z=p_{\pi}/p_{jet}$.} 
\end{figure}

\noindent Systematic errors were conservatively estimated to be $\pm0.023$, independent of $z$ and $j_T$,  for the $\pi^+$ and for the $\pi^-$ results.  The systematic error for the opposite charge sign pions should be treated as independent.  The dominant systematic error comes from limited statistical power in the simulation sample.  STAR simulations, composed of the PYTHIA 6.4 Monte Carlo generator\cite{Sjostrand:2006za} tuned to CDF Tune A and GEANT, were used to calculate the Collins moment at the particle and detector level.   The difference between the particle and detector level moments ($\delta{A}$) showed no dependence on $z$, $j_T$ or charge sign within statistical errors. The error on the mean of the $\delta{A}$ distribution was used as the systematic error.\\

\begin{figure}[h]
 \begin{minipage}{9.5cm}
  \includegraphics[height=.2\textheight]{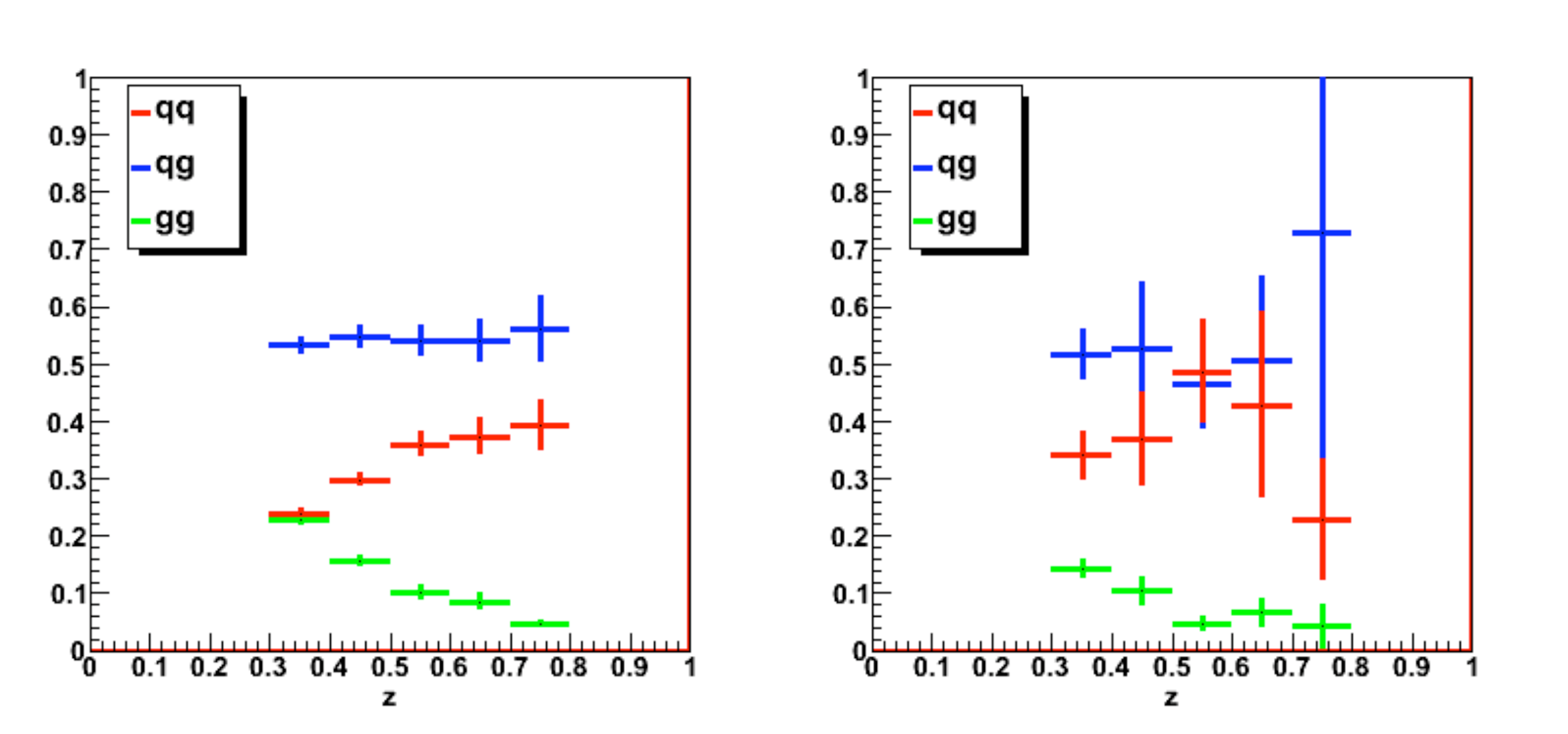}
  \hfill
\end{minipage}
\hfill
\begin{minipage}{7.0cm}
  \includegraphics[height=.18\textheight]{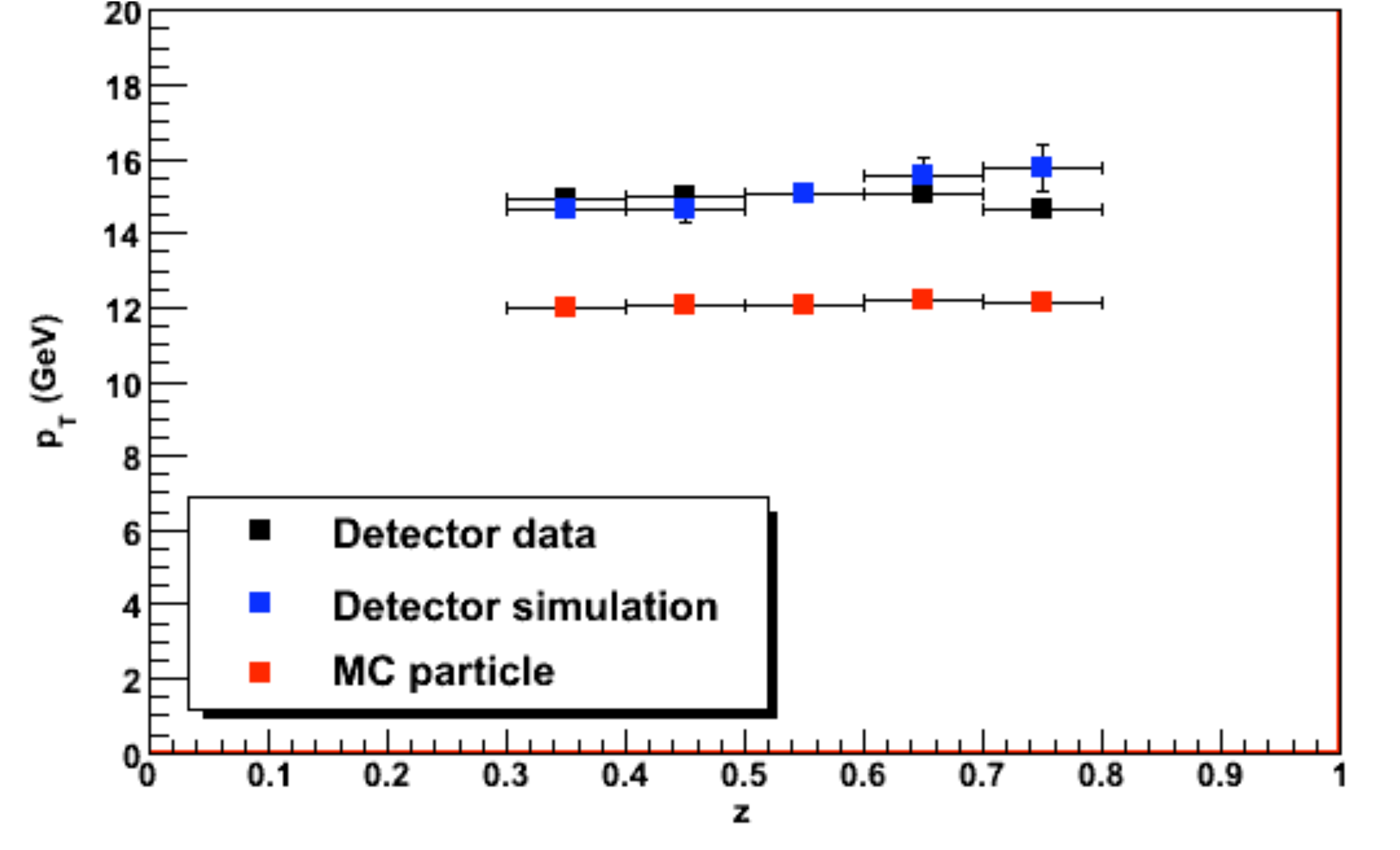}
\end{minipage}
\hfill   
\caption{The left panel shows the qq, qg and gg subprocess fractions at the particle level in simulation.  The middle panel demonstrates the change in the subprocess fractions introduced by applying trigger effects at the detector level in the same simulation.  The right panel shows that the enhancement of qq scattering in the data sample is due to the increase in the average jet $p_T$ at the detector level compared to the particle level.  The red points are the average jet $p_T$  at the particle level while the blue and black points are for the detector level in simulation and data respectively.}
 \label{fig:triggerbias}
\end{figure}

\noindent The second largest systematic error is due to trigger bias and is an order of magnitude smaller  than the dominant systematic error.  The right side panel in figure \ref{fig:triggerbias} shows that the trigger condition increases the average jet $p_T$ within a measured $z$ bin.  The fraction of qq (gg) scattering events increases (decreases) with increasing jet $p_T$ so a higher average jet $p_T$ translates directly into an enhancement of quark jets in the data sample.  Although this bias works in favor of an analysis that aims to measure the quark $\Delta_T{f(x)}$, it will also change the theoretical expectation for the asymmetry.  In order to facilitate comparisons to theory,  the change in subprocess fraction from the particle to detector level was determined using simulations as shown in the left and middle panels in figure \ref{fig:triggerbias}.  The change in subprocess fraction propagates to an increase (decrease) of $2.5\times10^{-3}$  in the measured  $\pi^+ (\pi^-)$ moments.   The systematic error due to detector resolution is similar in size to the trigger systematic, but moves the measurement in the opposite direction.  Detector resolutions will smear out the signal, reducing it in magnitude by an estimated 1.7$\times10^{-3}$.  Contamination from kaons, electrons and positrons contribute $1.2\times10^{-3}$ to the total error budget.  The uncertainty on the beam polarization is 4.8$\%$. All other systematic effects are estimated to be $<1\times10^{-3}$.  \\

\section{Conclusion}

\begin{figure}[h]
 \label{fig:results}
  \includegraphics[height=.3\textheight]{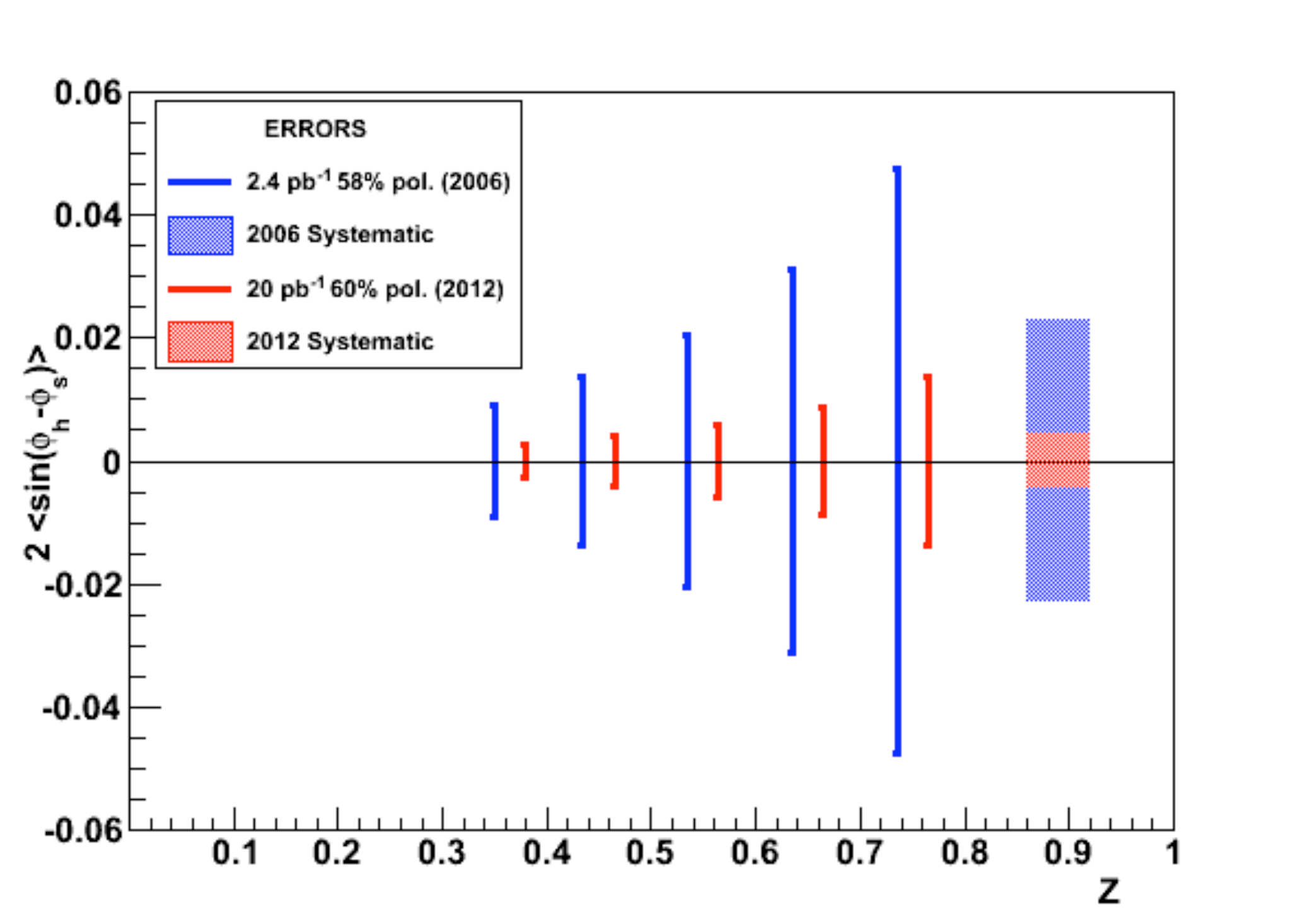}
  \caption{The current (blue) and projected (red) statistical and systematic error bars for this analysis.  } 
  \label{fig:projections}
\end{figure}

This proceeding presents the first analysis of the azimuthal distribution of leading charged pions within mid-rapdity jets reconstructed with the STAR detector.  The result is limited by systematic errors at the lower $z$ region and by statistical errors for $z$ > 0.6.   These errors limit the significance of the charge sign separation of the asymmetry to a $\sim1\sigma$ effect.  The authors anticipate a significant reduction in these systematic error bars before publication as well as in the analysis of future data.  Figure \ref{fig:projections} indicates the expected reduction in statistical and systematic errors if STAR collects the expected 20 $pb^{-1}$ during $\sqrt{s}=200$ GeV transverse spin running in the 2012 RHIC run.  The significant reduction in error is due to the increased figure of merit, improved trigger algorithms and increased bandwidth due to data acquisition upgrades.\\

\noindent This first measurement, while of limited statistical power, has laid the groundwork for future analysis of the Collins-like and Sivers-like moments in mid-rapidity jets as described in \cite{D'Alesio:2010am}.  Data from hadronic collisions provides complementary information to the existing SIDIS measurements, first by extending the kinematic reach of the measurements, and even more importantly by testing the process dependence and universality of the transverse momentum dependent distributions.

\begin{theacknowledgments}
This material is based in part upon work supported by the National Science Foundation under Grant No. 0855498. 
 \end{theacknowledgments}

\bibliographystyle{unsrtnat}
\bibliography{Fatemi_PANIC2}
\end{document}